# Small-world communication of residues and significance for protein dynamics

**Ali Rana Atilgan,**[1] **Pelin Akan,**[2] **and Canan Baysal**[2]

[1]School of Engineering, Bogazici University, Bebek 34342, Istanbul, Turkey.

[2]Laboratory of Computational Biology, Faculty of Engineering and Natural Sciences, Sabanci University, Orhanli 34956, Tuzla, Istanbul, Turkey.

**Correspondence**:

Canan Baysal

Phone: +90 (216) 483 9523

Fax: +90 (216) 483 9550

e-mail: canan@sabanciuniv.edu

**ABSTRACT**

It is not merely the position of residues that are of utmost importance in protein function and stability, but the interactions between them. We illustrate, by using a network construction on a set of 595 non-homologous proteins, that regular packing is preserved in short-range interactions, but short average path lengths are achieved through some long-range contacts. Thus, lying between the two extremes of regularity and randomness, residues in folded proteins are distributed according to a "small-world" topology. Using this topology, we show that the core residues have the same local packing arrangements irrespective of protein size. Furthermore, we find that the average shortest path lengths are highly correlated with residue fluctuations, providing a link between the spatial arrangement of the residues and protein dynamics.

**INTRODUCTION**

Proteins are tolerant to mutations with their liquid-like free volume distributions (Baase et al. 1999); however, the average packing density in a protein is comparable to that inside crystalline solids (Tsai et al. 2000). It has been shown that the interior of proteins are more like randomly packed spheres near their percolation threshold and that larger proteins are packed more loosely than smaller proteins (Liang and Dill 2001).

At physiological temperatures, the conformational flexibility is essential for biological activity that requires a concerted action of residues located at different regions of the protein (Baysal and Atilgan 2002; Zaccai 2000). This cooperation requires an infrastructure that permits a plethora of fast communication protocols. Highly transitive local packing arrangements, giving rise to regular packing geometries (Raghunathan and Jernigan 1997) cannot provide such short distances between highly separated residues for fast information sharing. On average, random packing of hard spheres similar to soft condensed matter are obtained for a set of representative proteins (Soyer et al. 2000). This architecture is capable of organizing short average path lengths between any two nodes in a structure, but it cannot warrant a high clustering similar to regular packing.

A network is referred to as a small-world network (SWN) if the average shortest path between any two vertices scales logarithmically with the total number of vertices, provided that a high local clustering is observed (Watts and Strogatz 1998). The former property of short paths is responsible for the name "small world." Neither regular configurations nor random orientations seem to exhibit these two intrinsic properties that are common in real-world complex networks (Newman 2000; Strogatz 2001). Proteins function efficiently, accurately and rapidly in the crowded environment of the cell; to this end, they should be effective information transmitters by design. With their ordered secondary structural units made up of α-helices and β-sheets on the one hand, and their seemingly unstructured loops on the other, proteins may well have the SWN organization (Vendruscolo et al. 2002).

In this study, we treat proteins as networks of interacting amino acid pairs (Atilgan et al. 2001; Bahar et al. 1997; Yilmaz and Atilgan 2000). We term these networks as "residue networks" to distinguish them from "protein networks" which are used to describe systems of interacting proteins (Jeong et al. 2001). We carry out a statistical analysis to show that proteins may be treated within the SWN topology. We analyze the local and global properties of these networks with their spatial location in the three dimensional structure of the protein. We also show that the shortest path lengths in the residue networks and residue fluctuations are highly correlated.

**METHODS**

**Spatial residue networks.** We utilize 595 proteins with sequence homology less than 25 % (Casadio 1999). We form spatial residue networks from each of these proteins using their Cartesian coordinates reported in the protein data bank (PDB) (Berman et al. 2000). In these networks, each residue is represented as a single point, centered on either the $C_\alpha$ or $C_\beta$ atoms; in the latter case, $C_\alpha$ atoms are used for Glycine residues. Since the general findings of this study are the same irrespective of this choice, we report results from the networks formed of $C_\beta$'s for brevity. Given the $C_\beta$ coordinates of a protein with $N$ residues, a contact map can be formed for a selected cut-off radius, $r_c$, an upper limit for the separation between two residues in contact. This contact map also describes a network which is generated such that if two residues are in contact, than there is a connection (edge) between these two residues (nodes) (Atilgan et al. 2001; Bahar et al. 1997; Yilmaz and Atilgan 2000). An example network formed for the protein 1ice is shown in figure 1. Thus, the elements of the so-called adjacency matrix, **A**, are given by

$$A_{ij} = \begin{cases} H(r_c - r_{ij}) & i \neq j \\ 0 & i = j \end{cases} \quad (1)$$

Here, $r_{ij}$ is the distance between the $i$th and $j$th nodes, H(x) is the Heavyside step function given by H(x) = 1 for x > 0 and H(x) = 0 for x ≤ 0.

**Network parameters.** The networks are quantified by local and global parameters, all of which can be derived from the adjacency matrix. The connectivity $k_i$ of residue $i$, is the number of neighbors of that residue:

$$k_i = \frac{1}{N} \sum_{j=1}^{N} A_{ij} \quad (2)$$

The average connectivity of the network is thus $K = <k_i>$, where the brackets denote the average.

The characteristic path length, $L$, of a network is the average over the minimum number of connections that must be transversed to connect residue pairs $i$ and $j$. In computing the shortest path between pairs of nodes, we make use of the fact that the number of different paths connecting a pair of nodes $i$ and $j$ in $n$ steps is given by, $B_{ij} = (\mathbf{A}^n)_{ij}$. Thus, the shortest path between nodes $i$ and $j$, $L_{ij}$, is given by the minimum power, $m$, of **A** for which $(\mathbf{A}^m)_{ij}$ is non-zero. The characteristic path length of the network is the average,

$$L = \frac{2}{N(N-1)} \sum_{i=1}^{N-1} \sum_{j=i+1}^{N} L_{ij} \quad (3)$$

Note that $L$ is a measure of the global properties, reflecting the overall efficiency of the network.

The clustering coefficient, *C*, on the other hand, reflects the probability that the neighbors of a node are also neighbors of each other, and as such, it is a measure of the local order. For residue *i* this probability may be computed by

$$C_i = \frac{\frac{1}{2}\sum_{j=1}^{N}\sum_{k=1}^{N} A_{ij} A_{ik} A_{kj}}{_{k_i}C_2} \quad (4)$$

Here $_{k_i}C_2$ is the combination relationship, and $k_i$ is the connectivity as defined in equation 2. The clustering coefficient of the network is the average $C = <C_i>$.

**Random rewiring of the residue networks.** For comparison purposes, we also generate random networks. The property common to the actual residue network and its random variant is the contact number of a given residue at a fixed cut-off radius. We rewire every residue (node) randomly to another residue chosen from a uniform distribution such that (i) it has the same number of neighbors (i.e. $k_i$ and *K* are the same as the residue network, but *C* and *L* change); and (ii) the chain connectivity is preserved by keeping the (*i*, *i*+1) contacts intact for all cutoff distances, $r_c$.

**RESULTS**

Within the framework of a local interaction network, residues in proteins organize into a SWN topology (see the **Appendix** for details). Our aim is to study the network topology of residue interactions from a statistical perspective so as to reveal the role of local arrangement on the overall structure and the dynamics of proteins. In the rest of this study, we present the results from the residue networks that are constructed using a 7 Å cutoff distance; we have verified that the general conclusions of this work are not affected when a 8.5 Å cutoff distance is used instead.

**Connectivity distribution of residues is independent of their spatial location.** The connectivity distribution of self-organizing networks has been shown to have direct consequences on the relative weight of (i) optimal performance, and (ii) tolerance to disturbances of these networks (Newman et al. 2002). At the extreme, scale-free networks are optimal for very fast communication between various parts. They are also very robust towards uncertainties for which they were designed, but are highly vulnerable towards unanticipated perturbations (Carlson and Doyle 2000). On the other hand, networks may be designed to become more tolerant to attack at the expense of some efficiency, by the utility of broad-scale or single-scale connectivity (Newman et al. 2002). Therefore, connectivity distribution should also be an indicator of the efficiency of information transfer in proteins.

A plot of the connectivity distribution is displayed in figure 2 for the residue networks studied here. We verify that the connectivity distribution of the residue networks constructed at a cutoff distance of 7 Å, which corresponds to the location of the first coordination shell, conform to the Gaussian distribution with a mean of 6.9 Å. It has been suggested that one of the main reasons for deviations from a scale-free connectivity distribution is the limited capacity of a given node (Amaral et al. 2000). In residue networks, this would translate into the excluded volume effect, since the number of residues that can physically reside within a given radius is limited.

Globular proteins may be considered to be made up of a core region surrounded by a molten layer of surface residues. It is of interest to distinguish the topological differences between the core and the surface. Thus, we have also investigated the connectivity distribution of the core and surface residues. We utilize the DEPTH program which differentiates between such residues by calculating the depth of a residue from the protein surface (Chakravarty and Varadarajan 1999). We classify the core residues as those residing at depths larger than 4 Å, based on a previous study (Baysal and Atilgan 2002). We find that the same type of distribution of coordination numbers is valid for both the hydrophobic core and the molten surface, as shown by the separate contact distribution of the surface and core residues (figure 2). The means for the respective cases are 5.0 and 8.4 Å. This demonstrates that, the *same* small-world organization prevails throughout the protein, despite the heterogeneous density distribution.

**Clustering of residues is independent of their location in the core.** We have further investigated the shortest average path length $L_i$ and the clustering coefficient $C_i$ of residue $i$ as a function of residue depth $D_i$. For this purpose, we have again used residue depth as a measure of its location in the folded protein. To eliminate the size effect, we have studied a subset of proteins of a fixed number of residues. In figure 3, $L_i$ and $C_i$ as a function of residue depth is shown for proteins of size 150±10, 210±10, and 310±10; averages are taken over 24, 15, and 15 proteins in the respective cases.

As expected, the shortest path length decreases for residues at greater depths, i.e. those in the core of the protein are connected to the rest of the residues in a fewer number of steps; moreover, this property is size dependent as corroborated by the logarithmic size dependence of the characteristic path lengths (see figure 6 in the **Appendix**). Perhaps much less expected, on the other hand, is that the clustering coefficient approaches a fixed value of ca. 0.35 beyond a depth of ca. 4 Å irrespective of the size of the proteins studied. At greater depths, where the residues are completely surrounded by other residues and are not exposed to the solvent, the local organization of the protein is always the same.

**Shortest path lengths and fluctuations are related.** Residue fluctuations, which are both experimentally and computationally accessible, provide a rich source of information on the dynamics of proteins around their folded state. It is possible to discern the functionally important motions in proteins using a modal decomposition of the cross-correlations of the fluctuations (Bahar et al. 1998a). Fontana and collaborators have elegantly demonstrated that limited proteolysis, a biochemical method that can be used as a probe of structure and dynamics of both native and partly folded proteins, does not occur at just any site located on the protein surface, but rather shows a good correlation with larger crystallographic B-factors (see Tsai et al. 2002 and references cited therein). Some correlation has also been demonstrated between the residue fluctuations and the native state hydrogen exchange data of folded proteins, the latter providing information on the local conformational susceptibilities of residues (Bahar et al. 1998b).

Thus, repeatedly, residue fluctuations around the folded state emerge as a measurable that can be related to the dynamics of the protein. One would expect an indirect correlation between the fluctuations and shortest path lengths: The former are smaller for highly connected residues, which are in turn connected to the rest of the molecule, on average, in a shorter number of steps. Our analysis on numerous proteins has shown that residue fluctuations are also highly correlated with the shortest path lengths, $L_i$. In this study residue fluctuations are computed by the Gaussian Network Model of proteins,[1] which was shown to be in excellent agreement with crystallographic B-factors (Bahar et al. 1999; Baysal and Atilgan 2001; Ming et al. 2003).

---

[1] According to this model, average residue fluctuations are given by, $\langle \Delta R_i^2 \rangle \alpha \left[ \mathbf{\Gamma}^{-1} \right]_{ii}$. $\mathbf{\Gamma}$ is the Kirchoff matrix whose diagonal entries represent the packing density of the $i$th residue, and the off-diagonal elements are given by the negative of the adjacency matrix elements given by equation 1.

Example comparisons between the fluctuations and path lengths are displayed in figure 4 for α, β, α+β, and α/β proteins. Note that the correlation that emerges between the fluctuations and path lengths exceeds the expectations from the simple inference outlined above, based on connectivity arguments. Therefore, there is an intriguing balance between these two measurables, one of which ($L_i$) is more readily associated with the global features and the other (fluctuations) with the local features of the network.

**CONCLUDING REMARKS**

We have shown that the protein structure may be classified as a SWN, balancing efficiency and robustness. We find that the same local organization of core residues appears irrespective of the protein size. Moreover, a remarkable correlation exists between residue fluctuations and shortest path lengths. This unifying network perspective will let us explore protein dynamics such that, among other things, we will be able to (i) distinguish structurally important residues in folding, binding, and stability, (ii) locate the routes through which a perturbation is communicated in a protein, and (iii) estimate the time scales on which a response is generated. The spatio-temporal nature of the hypothesized process calls for deeper investigation on particular proteins. The global rules deduced here for proteins are also expected to have applications in bioinformatics problems such as identifying interaction surfaces in protein docking and distinguishing misfolded states.

**APPENDIX: Residues in proteins organize in a small-world-network topology.**

In SWNs, the measure of global communication between any two nodes, characterized by the characteristic path length, $L$, is on the same order of magnitude as a random network. At the same time, the local structure needs to be organized such that the probability that the neighbors of a node are also neighbors of each other is high; in a random network, such a construction does not exist. The latter property is quantified by the clustering coefficient, $C$ (Watts 1999), which is at least about one order of magnitude larger in SWNs than in their randomized counterparts (Watts and Strogatz 1998). The final condition for a small-world behavior in a network is that the average path length should scale logarithmically with the total number of vertices (Davidsen et al. 2002). These conditions are summarized as:

$$\begin{aligned} L &\approx L_{\text{random}} \ll L_{\text{regular}} \\ C &\gg C_{\text{random}} \\ L &\propto \log N \end{aligned} \quad (A1)$$

We first study the ratios $L/L_{\text{random}}$ and $C/C_{\text{random}}$ to understand if the first two of these conditions are met in residue networks. The results are presented in figure 5 as a function of the cutoff distance, $r_c$. We find that $L$ is on the same order as $L_{\text{random}}$ for all values of $r_c$ (right $y$-axis). For shorter distances ($r_c \leq 8.5$ Å) the average path length in real proteins is found to be ca. 1.8 times that of random networks; the ratio gradually decreases towards the theoretical limit of 1 as $r_c$ is increased. The clustering coefficient, $C$, of the residue networks, on the other hand, is ca. 9-13 times that of random ones for $r_c \leq 8.5$ Å. For larger $r_c$, the ratio rapidly falls to 1 (left $y$-axis).

The final condition of equation A1 for a small-world behavior in a network is that the average path length should scale logarithmically with the total number of vertices (Davidsen et al. 2002). Such a scaling is observed for the proteins studied in this work. A representative case for $r_c = 7$ Å is shown in figure 6. Note that in reproducing this figure, we have clustered the proteins used in this study according to size such that a point corresponding to protein size $N$ corresponds to an average over all proteins in our set that fall in the range $N \pm 10$. Also shown in this figure is the logarithmic scaling of the randomized counterparts of the residue networks. Note that the slope of the latter is $1/\log K$, a well known result for Poisson and Gaussian distributed random graphs (Newman et al. 2001).

Thus, interactions within proteins behave like SWNs in the cutoff distance range of up to ca. 8.5 Å. We note that Vendruscolo et al. have studied a set of 978 proteins at a cut-off distance of 8.5 Å with the network perspective. They find that $L$ is 4.1 ± 0.9 and $C$ is 0.58 ± 0.04; they do not show the logarithmic dependence of $L$ on system size, $N$ (last condition in equation A1). Nevertheless, based on the small value of the average path length and the relatively large value of the clustering coefficient, they conclude that native protein structures belong to the class of small-world graphs (Vendruscolo et al. 2002), a valid assertion for the 8.5 Å cutoff. To clarify the physical meaning of a cutoff distance in the context of network topology, we look at the radial distribution function for residues in proteins (inset to figure 5). Cutoff values of ca. 6.5 – 8.5 Å have been used in studies where coarse graining of proteins is utilized (Bagci et al. 2002; Dokholyan et al. 2002; Miyazawa and Jernigan 1996). The lower bound corresponds to the first coordination shell of the protein; i.e. the range within which residue pairs are found with the highest probability (6.7 Å for the set used here; first hump in the inset to figure 5). A great portion of the contribution to this shell is due to chain connectivity; all ($i$, $i$+1) and most ($i$, $i$+2) pairs fall within this range. Non-bonded residue pairs also exist in this coordination shell. However, the contribution of non-bonded pairs to higher order coordination shells may also be significant (Woodcock 1997). For $C_\beta - C_\beta$ interactions in proteins, the second shell occurs at 8.6 (the second hump in the inset to figure 5). Above, we have shown that residues in proteins form small-world networks for the first and second coordination shells. Beyond the second coordination shell the clustering coefficient, $C$, which is a local property, looses physical significance.

**ACKNOWLEDGEMENTS**

Partial support provided by DPT (Devlet Planlama Teskilati) Project Grant No. 01K120280 is acknowledged.

**REFERENCES**

Amaral LAN, Scala A, Barthelemy M, Stanley HE. 2000. Classes of Small-World Networks. Proc. Natl. Acad. Sci. USA 97:11149-11152.

Atilgan AR, Durell SR, Jernigan RL, Demirel MC, Keskin O, Bahar I. 2001. Anisotropy of Fluctuation Dynamics of Proteins with an Elastic Network Model. Biophys. J. 80:505-515.

Baase WA, Gassner NC, Zhang X-J, Kuroki R, Weaver LH, Tronrud DE, Matthews BW. 1999. How Much Sequence Variation Can the Functions of Biological Molecules Tolerate. In: Frauenfelder H, Deisenhofer J, Wolynes PG, editors. Simplicity and Complexity in Proteins and Nucleic Acids: Dahlem University Press.

Bagci Z, Jernigan RL, Bahar I. 2002. Residue Packing in Proteins: Uniform Distribution on a Coarse-Grained Scale. J. Chem. Phys. 116:2269-2276.

Bahar I, Atilgan AR, Demirel MC, Erman B. 1998a. Vibrational Dynamics of Folded Proteins: Significance of Slow and Fast Modes in Relation to Function and Stability. Phys. Rev. Lett. 80:2733-2736.

Bahar I, Atilgan AR, Erman B. 1997. Direct Evaluation of Thermal Fluctuations in Proteins Using a Single Parameter Harmonic Potential. Fold. Des. 2(3):173-181.

Bahar I, Erman B, Jernigan RL, Atilgan AR, Covell DG. 1999. Collective Dynamics of Hiv-1 Reverse Transcriptase: Examination of Flexibility and Enzyme Function. J. Mol. Biol. 285:1023-1037.

Bahar I, Wallqvist A, Covell DG, Jernigan RL. 1998b. Correlation between Native-State Hydrogen Exchange and Cooperative Residue Fluctuations from a Simple Model. Biochemistry 37:1067-1075.

Baysal C, Atilgan AR. 2001. Elucidating the Structural Mechanisms for Biological Activity of the Chemokine Family. Proteins 43:150-160.

Baysal C, Atilgan AR. 2002. Relaxation Kinetics and the Glassiness of Proteins: The Case of Bovine Pancreatic Trypsin Inhibitor. Biophys. J. 83:699-705.

Berman HM, Westbrook J, Feng Z, Gilliland G, Bhat TN, Weissig H, Shindyalov IN, Bourne PE. 2000. The Protein Data Bank. Nucl. Acids Res. 28:235-242.

Carlson JM, Doyle J. 2000. Highly Optimized Tolerance: Robustness and Design in Complex Systems. Phys. Rev. Lett. 84:2529-2532.

Casadio PFaR. 1999. A Neural Network Based Predictor of Residue Contacts in Proteins. Prot. Eng. 12:15-21.

Chakravarty S, Varadarajan R. 1999. Residue Depth: A Novel Parameter for the Analysis of Protein Structure and Stability. Structure 7:723-732.

Davidsen J, Ebel H, Bornholdt S. 2002. Emergence of a Small World from Local Interactions: Modeling Acquaintance Networks. Phys. Rev. Lett. 88:art. no. 128701.

Dokholyan NV, Li L, Ding F, Shakhnovich EI. 2002. Topological Determinants of Protein Folding. Proc. Natl. Acad. Sci. USA 99:8637-8641.

Jeong H, Mason SP, Barabasi A-L, Oltvai ZN. 2001. Lethality and Centrality in Protein Networks. Nature 411:41-42.

Liang J, Dill KA. 2001. Are Proteins Well Packed? Biophys. J. 81:751-766.

Ming D, Kong Y, Wu Y, Ma J. 2003. Substructure Synthesis Method for Simulation Large Molecular Complexes. Proc. Natl. Acad. Sci. USA 100:104-109.

Miyazawa S, Jernigan RL. 1996. Residue-Residue Potentials with a Favorable Contact Pair Term and an Unfavorable High Packing Density Term, for Simulation and Threading. J. Mol. Biol. 256:623-644.

Newman MEJ. 2000. Models of the Small World. J. Stat. Phys. 101:819-841.

Newman MEJ, Girvan M, Farmer JD. 2002. Optimal Design, Robustness, and Risk Aversion. Phys. Rev. Lett. 89:art. no. 028301.

Newman MEJ, Strogatz SH, Watts DJ. 2001. Random Graphs with Arbitrary Degree Distributions and Their Applications. Phys. Rev. E 64:art. no. 026118.

Raghunathan G, Jernigan R. 1997. Ideal Architecture of Residue Packing and Its Observation in Protein Structures. Prot. Sci. 6:2072-2083.

Soyer A, Chomilier J, Mornon J-P, Jullien R, Sadoc J-F. 2000. Voronoi Tessellation Reveals the Condensed Matter Character of Folded Proteins. Phys. Rev. Lett. 85:3532-3535.

Strogatz SH. 2001. Exploring Complex Networks. Nature 410:268-276.

Tsai AM, Neumann DA, Bell LN. 2000. Molecular Dynamics of Solid-State Lysozyme as Affected by Glycerol and Water: A Neutron Scattering Study. Biophys. J. 79:2728-2732.

Tsai C-J, de Laureto PP, Fontana A, Nussinov R. 2002. Comparison of Protein Fragments Identified by Limited Proteolysis and by Computational Cutting of Proteins. Prot. Sci. 11:1753-1770.

Vendruscolo M, Dokholyan NV, Paci E, Karplus M. 2002. Small-World View of the Amino Acids That Play a Key Role in Protein Folding. Phys. Rev. E 65:art. no. 061910.

Watts DJ. 1999. Small Worlds. Princeton: Princeton University Press.

Watts DJ, Strogatz SH. 1998. Collective Dynamics of 'Small-World' Networks. Nature 393:440-442.

Woodcock LV. 1997. Entropy Difference between the Face-Centered Cubic and Hexagonal Close-Packed Structures. Nature 385:141-143.

Yilmaz LS, Atilgan AR. 2000. Identifying the Adaptive Mechanism in Globular Proteins: Fluctuations in Densely Packed Regions Manipulate Flexible Parts. J. Chem. Phys. 113(10):4454-4464.

Zaccai G. 2000. How Soft Is a Protein? A Protein Dynamics Force Constant Measured by Neutron Scattering. Science 288:1604-1607.

**FIGURE CAPTIONS:**

*Figure 1.* Network construction from a protein. Here the structure of human interleukin 1-β converting enzyme (PDB code: 1ice) is shown on the left. The network constructed from the $C^{\beta}$ coordinates of the residues ($C^{\alpha}$ for Gly) at 7 Å cutoff is shown on the right.

*Figure 2.* Residue contact distribution at $r_c$ = 7 Å, computed as an average over all the residues in a set of 54 proteins. The familiar form of the contact distribution is captured (see, for example, figure 4 in Miyazawa and Jernigan). The contact distributions of core and surface residues are also displayed. Gaussian distribution of coordination numbers is valid for both the hydrophobic core and the molten surface.

*Figure 3.* The depth dependence of the characteristic path length (empty symbols) and the clustering coefficient (filled symbols) for proteins of fixed sizes ($N$=150: squares, 24 proteins; $N$=210: triangles, 15 proteins; $N$=310: circles, 15 proteins). The characteristic path length consistently decreases for residues at greater depths; moreover, its value depends on system size. On the other hand, at depths greater than 4 Å, the clustering coefficient attains a fixed value of ca. 0.35 irrespective of system size and the location of the residue. Even for the surface residues, the clustering coefficient is independent of system size, although its value is location dependent and somewhat higher than 0.34.

*Figure 4.* A good correlation between the shortest path lengths and residue fluctuations is observed. Four examples, one of each from α, β, α+β, and α/β class of proteins, are displayed

*Figure 5.* In a SWN, characteristic path length, $L$, is on the same order of magnitude as its randomized counterpart, whereas clustering density, $C$, is at least one order of magnitude larger. The variation of the ratios $L/L_{random}$ (right ordinate) and $C/C_{random}$ (left ordinate) in the residue networks with the cut-off distance, $r_c$, used in forming the networks is shown. Note that as $r_c \rightarrow \infty$ both $L$ and $C$ approach 1, since every node will be connected to every other node at this limit. **Inset:** Radial distribution function of the residue networks. All data are averages over 595 nonhomologous proteins.

*Figure 6.* In a SWN, the characteristic path length, $L$, should show a logarithmic dependence on the system size, $N$. Thus, the relation $L \propto \log(N)$ should hold up to a cutoff value of ca. 8.5 Å. An example case for $r_c$ = 7 Å is shown. Also shown is the logarithmic dependence of $L$ on $N$ for the randomized networks, the slope of which is the inverse logarithm of the connectivity. That the relationship $L = \log N / \log K$ should hold for Poisson and Gaussian distributed random networks is a well known result (Newman et al. 2001).

Figure 1

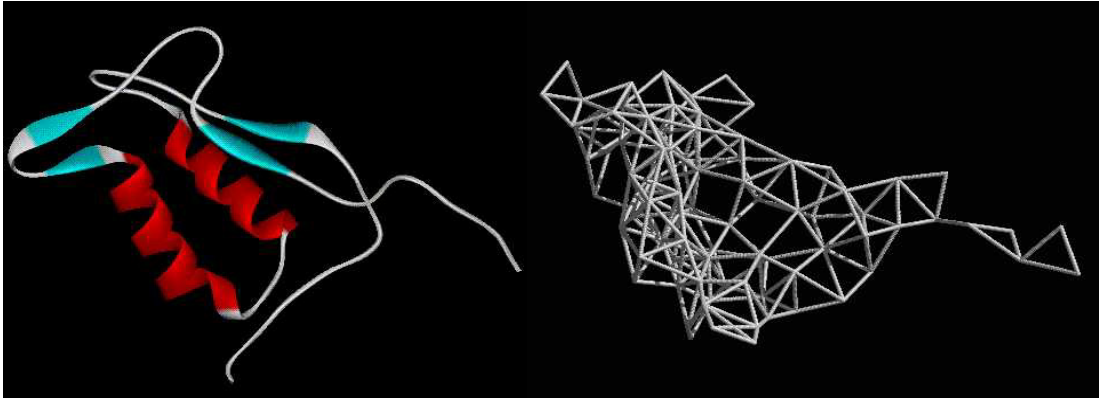

Figure 2

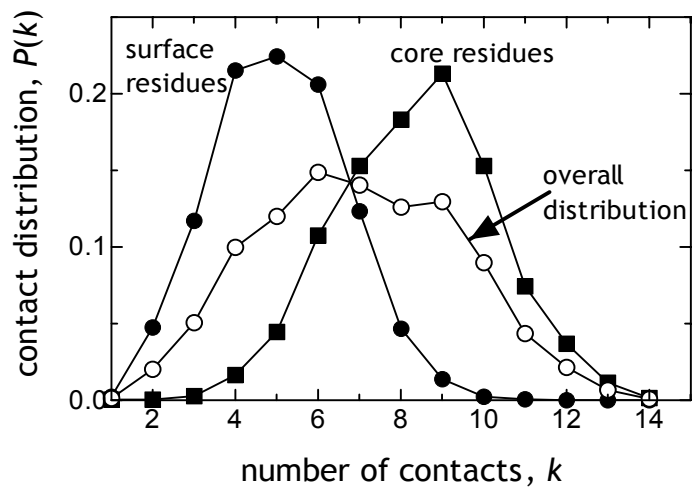

Figure 3

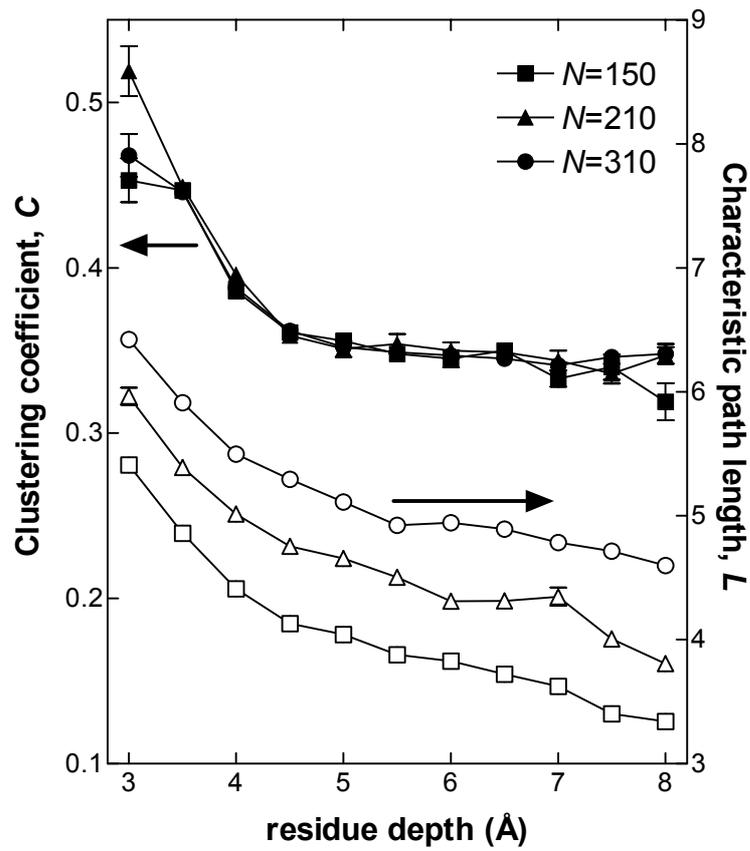

Figure 4

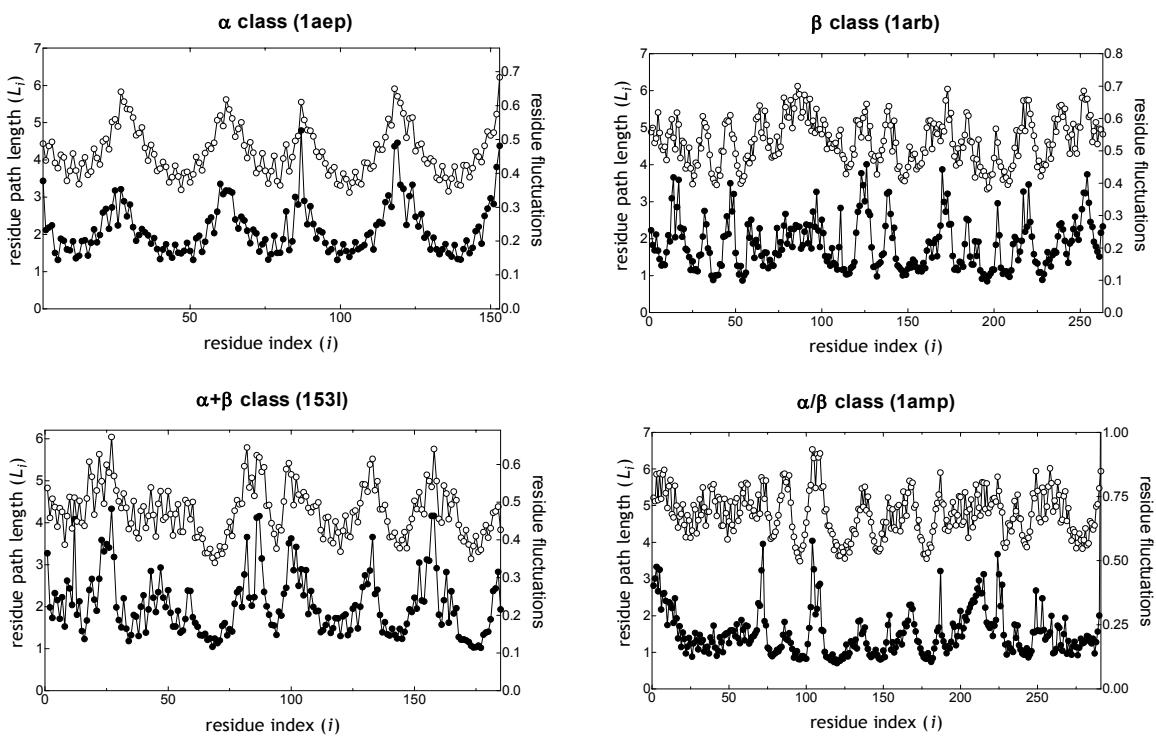

Figure 5

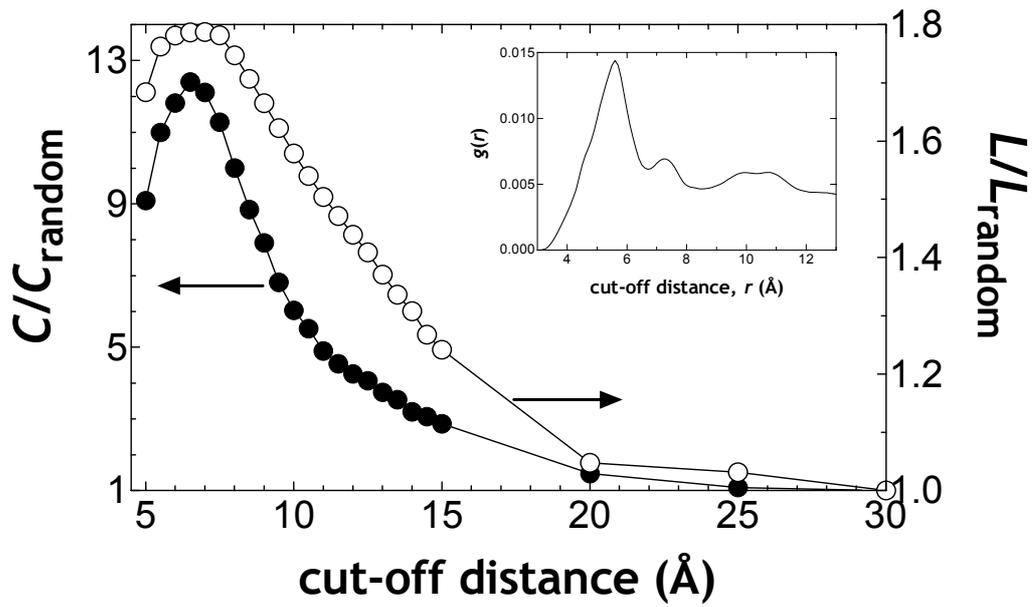

Figure 6

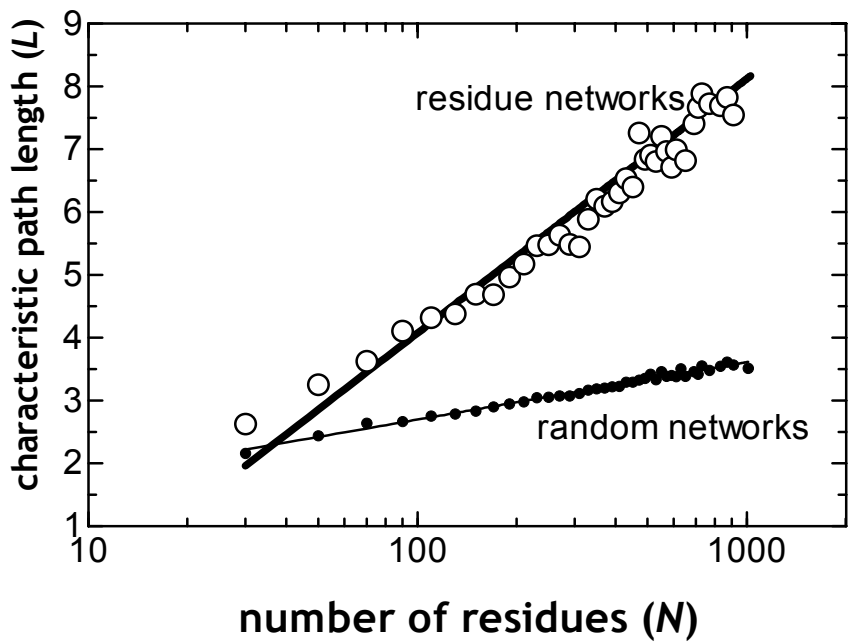